\documentclass[aps,pra,twocolumn,groupaddress,showpacs,floatfix]{revtex4}
\usepackage{graphicx}
\usepackage{dcolumn}
\usepackage{bm}
\usepackage{amsfonts}
\usepackage{amsmath}

\begin{document}
\baselineskip=0.5cm
\renewcommand{\thefigure}{\arabic{figure}}

\title{Spin density-functional theory for imbalanced interacting Fermi
gases in highly elongated harmonic traps}

\author{Gao Xianlong}
\affiliation{Department of Physics, Zhejiang Normal University,
Jinhua, Zhejiang Province, 321004, China}
\author{Reza Asgari}
\affiliation{School of Physics, Institute for Studies in Theoretical Physics and Mathematics, 19395-5531 Tehran, Iran }

\date{\today}

\begin{abstract}
We numerically study imbalanced two component Fermi gases with attractive interactions in highly elongated harmonic traps.
An accurate parametrization formula for the ground state energy is presented for a spin-polarized attractive Gaudin-Yang model.
Our studies are based on an accurate microscopic spin-density-functional theory through the
Kohn-Sham scheme which employs the one-dimensional homogeneous Gaudin-Yang model with Luther-Emery-liquid ground-state correlation as a
reference system.
A Thomas-Fermi approximation is examined incorporating the exchange-correlation interaction.
By studying the charge and spin density profiles of the
system based on these methods, we gain a quantitative understanding of the role of attractive
interactions and polarization on the formation of the two-shell structure,
with the coexisted Fulde-Ferrell-Larkin-Ovchinnikov-type phase in the center of the trap and either the BCS
superfluid phase or the normal phase at the edges of the trap.
Our results are in good agreement with the recent theoretical consequences.
\end{abstract}
\maketitle

\section{Introduction}
\label{sect:intro}

Progress made in trapping ultracold atomic gases into different
low-dimensional systems~\cite{moritz_2005,Paredes} has provided us
grounds to describe and simulate the behavior of fascinating
non-Fermi liquid in which the
analysis may not be as tractable. The physical properties of such
systems have attracted both experimental and theoretical interest.

For the research in the experimental side, the Tonks-Girardeau regime of
strongly repulsive bosons in one-dimensional (1D) systems has been
observed~\cite{Paredes}. This regime is understood through the mapping of the
strongly repulsive, impenetrable bosons onto an ideal gas of
fermions subjected to the same external potential~\cite{Girardeau}.
Esslinger and colleagues at ETH Zurich in Switzerland confined a
two-component Fermi gas of $^{40}$K atoms to thousands of highly
elongated one-dimensional tubes. Such a system
is realized by using a two-dimensional optical lattice, having about 100 atoms in each tube~\cite{moritz_2005}
and serves as a 1D matter waveguide to observe the two-particle bound states of atoms.
For the first time, they created confinement induced molecules.
The binding energy of the molecules is measured by using radio-frequency
spectroscopy bearing a good agreement with the theoretical
predictions~\cite{olshanii_1998}. Importantly, two-particle bound molecular states are formed in 1D
confined system independent of the nature of the interaction, i.e.,
attractive or repulsive between atoms. This is obviously different from what happens in free
space where a molecule is formed only when the atoms attract.

On the other hand, many theoretical efforts on quasi-one dimensional
(Q1D) inhomogeneous Fermi gases have been made. For 1D trapped
fermions at finite temperature with attractive binary interactions,
the transition to a fermion-paired state is obtained by exact stochastic
mean-field calculations~\cite{Juillet}, which is
characterized by dominant
Cooper pairing correlations, an algebraic long-range order and a superfluid component.
A one-dimensional spin-polarized Fermi gas with infinitely strong attractive
zero-range odd-wave interactions has been studied. This system is known as the
fermionic Tonks-Girardeau gas, arising from a confinement-induced
resonance reachable via a three-dimensional p-wave Feshbach
resonance~\cite{Minguzzi}. In Ref.~\cite{gao_2002}, a bosonization
technique has been developed to calculate analytically the density
profile, the momentum distribution, and several correlation
functions of two-component Fermi gases with inclusion of
forward-scattering processes and exclusion of backward scattering
between inter-components. Inhomogeneous Tomonaga-Luttinger liquid
model, with space-dependent parameters assuming a slowly varying
trap potential on the scale of the Fermi wavelength, has been used
to describe spin-charge separation in two-component Fermi gases~\cite{recati_2003,kecke_2005}. The Thomas-Fermi approximation is
widely used to calculate the ground state properties of a large
system~\cite{astrakharchik_2004,kecke_2005}. The existence of a
molecular Tonks-Girardeau gas is predicted for strong attractive
effective interactions~\cite{astrakharchik_2004}. In
Ref.~\cite{Gao_PRA06} a microscopic calculation of the
density-functional theory (DFT), based on the exact Bethe-ansatz
solution of 1D homogeneous Gaudin-Yang system, is presented for the
ground-state density profile with arbitrary size. Quantitative
understanding about the role of the repulsive or attractive
interactions on the shell structure of the axial density profile is
achieved.

Above mentioned studies are focused on the fully polarized Fermi
atomic gases. Recently, progress in two
experiments~\cite{Zwerlein,Partridge} in trapping partially
polarized Fermi atomic gases has paved the way for systematic
studies on systems with different number of spin-up and spin-down atoms ($N_\uparrow \ne N_\downarrow$).
The experiment explored on three-dimensional polarized $^6$Li
gases~\cite{Zwerlein} suggested that the density profiles show
coexisting three-shell structures: a fully unpolarized paired
superfluid phase in the center of the trap, a fully polarized
noninteracting phase composed of excess atoms, and a partially
polarized normal shell between these two regions. The experiment
which employed an elongated cigar-shaped trap of polarized $^6$Li gases~\cite{Partridge} suggests the gas separates into a phase that is
consistent with a superfluid paired core surrounded by a shell of
normal unpaired fermions beyond a critical polarization.

Recently two theoretical investigations~\cite{Orso,Hu} discussed
partially polarized two-component Fermi gases of attractive
interaction within local density approximation while implementing the
exact exchange-correlation interaction from the homogeneous
Bethe-ansatz solution of 1D Gaudin-Yang model. Both of these two papers predict a
two-shell density profile structure of a coexisting partially
polarized superfluid core in the center of the trap while at the
edges of the trap either fully paired or fully polarized wings
depending on the attractive interaction strength and the
polarization. Furthermore, they anticipated a critical spin polarization and a non-monotonic
behavior of the Thomas-Fermi radius of each spin component
as a function of the polarization. Orso~\cite{Orso}
and Hu et al.~\cite{Hu} identified the polarized superfluid as having
the Fulde-Ferrell-Larkin-Ovchinnikov (FFLO) structure, which is stabilized in a polarized
two-component Fermi gas in an array of weakly-coupled 1D tubes~\cite{Parish}. These two
theoretical efforts used the local-density approximation for the
harmonic trap and Thomas-Fermi-type approximation for the
interaction. The local-density approximation is applicable when
$N\gg 1$, where $N$ is the number of particles, but deteriorates with decreasing particle number and
completely neglects the tunneling of the density profile beyond the
Thomas-Fermi radius where one instead needs to use a method
beyond the local-density approximation, like the microscopic density-functional theory.

The FFLO phenomena of imbalanced fermion populations in 1D optical lattices
of attractive interactions have also been
extensively discussed recently~\cite{Tezuka,Feiguin,Batrouni},
where FFLO pairing is formed robust from the attractive
interaction for a wide range of polarization, and for the spatial inhomogeneities induced
by the presence of a trap in the experimental reality.

In this paper, we study 1D
harmonically trapped Fermi gases of imbalanced spin population with
two alternative approaches. One is the spin-density-functional theory based on the
Bethe-ansatz results for the 1D Luttinger liquid, which is specially
useful for systems whose analytical solutions or exact numerical
methods are not available, and the other is the Thomas-Fermi approximation based on the
non-interacting approximation for the kinetic energy. Both methods consider exactly the
exchange-correlation energy of the corresponding 1D homogeneous
system. Within these two methods, we quantitatively study the systems of weak and
intermediate coupling strength and different polarizations by analyzing
the charge and spin density profiles.

The paper is organized as follows. In Sec. II we introduce our
model and present the parametrization result for the ground state
energy of the homogeneous Gaudin-Yang model. In Sec. III the
spin-density-functional theory is briefly discussed. Numerical results
and discussion are presented in Sec. IV. Finally, we conclude in Sec. V.

\section{The model and its parametrization result}
\label{Section:II}
A system of $N$ spin-$1/2$ fermions is considered to be one-dimensional when
subjected to a strongly anisotropic harmonic potential with angular frequencies in the radial direction $\omega_\perp$ much larger
than that in the axial direction $\omega_\|$ with $\omega_{\|}/\omega_\perp\ll N^{-1}$.
We consider two kind of fermion particles with mass $m$,
interacting {\it via} a $\delta$-type contact with an effective 1D coupling strength $g_{\rm 1D}$
in the harmonic trap. The Hamiltonian for such a system is given by
\begin{eqnarray}\label{eq:igy}
{\cal H}&=&-\frac{\hbar^2}{2m}\sum_{i=1}^{N}\frac{\partial^2}{\partial z^2_i}
+g_{\rm 1D}\sum_{i=1}^{N_{\uparrow}}\sum_{j=1}^{N_{\downarrow}}\delta(z_i-z_j)
\nonumber\\
&&+\frac{1}{2}m\omega^2_{\|}\sum_{i=1}^{N}z^2_i\, .
\end{eqnarray}
Note that when p-wave interactions are neglected, the contact s-wave interactions between same species are suppressed by the Pauli exclusion principle.
The coupling constant $g_{1D}$ can be written in terms of the scattering strength as $g_{1D} =-2\hbar^2/(a_{1D}m)$.
The effective 1D scattering length $a_{1D}$ can be expressed
through the three-dimensional scattering length $a_{3D}$ for fermions confined
in a quasi-1D geometry, $a_{1D}=-a^2_\perp(1-Aa_{3D}/a_{\|})/a_{3D}>0$
with the constant $A\approx 1.0326$~\cite{olshanii_1998}. In the following,
we will choose the harmonic-oscillator length $a_{\|}=\sqrt{\hbar/(m\omega_{\|})}$
as unit of length and the harmonic-oscillator quantum $\hbar\omega_{\|}$ as unit of energy.

Without the external potential the Hamiltonian (\ref{eq:igy}) reduces to the homogeneous Gaudin-Yang model,
which is exactly solvable by use
of the Bethe-ansatz method~\cite{GY}. In the thermodynamic limit, the system is determined by
the spin polarization $\zeta=(N_\uparrow-N_\downarrow)/N$ and a dimensionless parameter $\gamma \equiv mg_{\rm 1D}/(\hbar^2 n)$, while
in the fully unpolarized system the only parameter of the system is $\gamma$.
We denote the charge density as $n=N/L$ and the spin density
$s=(N_\uparrow-N_\downarrow)/2L=(n_\uparrow-n_\downarrow)/2$.

For $N$ attractive fermions, the ground state is described by the coupled integral equations for the momentum distribution $\rho(k)$
and $\sigma(\lambda)$,
\begin{equation}\label{eq:rho}
\rho(k)=\frac{1}{2\pi}+\frac{2\gamma n}{\pi}\int^{B}_{-B}d\lambda
\frac{1}{(\gamma n)^2+4(k-\lambda)^2}\sigma(\lambda)\,,
\end{equation}
and  
\begin{eqnarray}\label{eq:sigma}
\sigma(\lambda)&=&\frac{1}{\pi}
+\frac{2\gamma n}{\pi}\int^{Q}_{-Q}d k \,\frac{1}{(\gamma n)^2+4(k-\lambda)^2}\rho(k)
\nonumber\\
&&+\frac{\gamma n}{\pi}\int^{B}_{-B}d\lambda'
\frac{1}{(\gamma n)^2+(\lambda-\lambda')^2}\sigma(\lambda')\,,
\end{eqnarray}
where $B$ and $Q$ are non-negative numbers related to the normalization conditions
\begin{equation}\label{eq:normalization_1}
\int^{Q}_{-Q}dk \rho(k)=2s
\end{equation}
and
\begin{equation}\label{eq:normalization_2}
\int^{B}_{-B}d\lambda \sigma(\lambda)=\frac{n}{2}-s\,.
\end{equation}
The ground state energy (GSE) per particle is written in terms of the momentum distributions $\rho(k)$ and $\sigma(\lambda)$ as
\begin{eqnarray}\label{eq:energy_atom}
\frac{E}{N}&=&
\frac{1}{n}\frac{\hbar^2}{2m}\left[\int^{Q}_{-Q}d k k^2\rho(k)\right.\nonumber\\
&&\left.+2\int^{B}_{-B}d\lambda \,\,[\lambda^2-(\gamma n/2)^2]\sigma(\lambda)\right]\,.
\end{eqnarray}
\begin{figure}
\begin{center}
\tabcolsep=0 cm
\includegraphics[width=0.65\linewidth,angle=0]{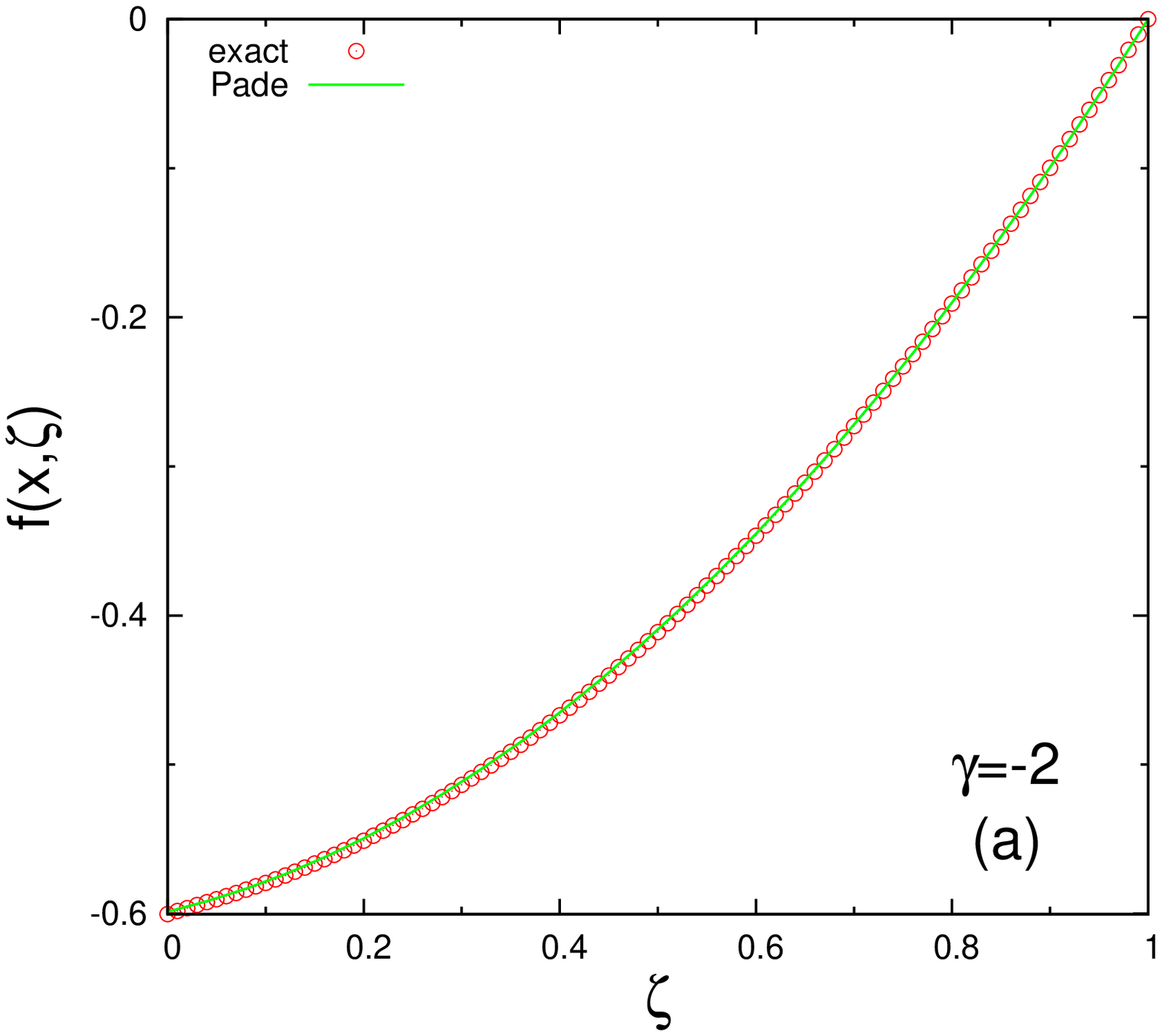}
\includegraphics[width=0.65\linewidth,angle=0]{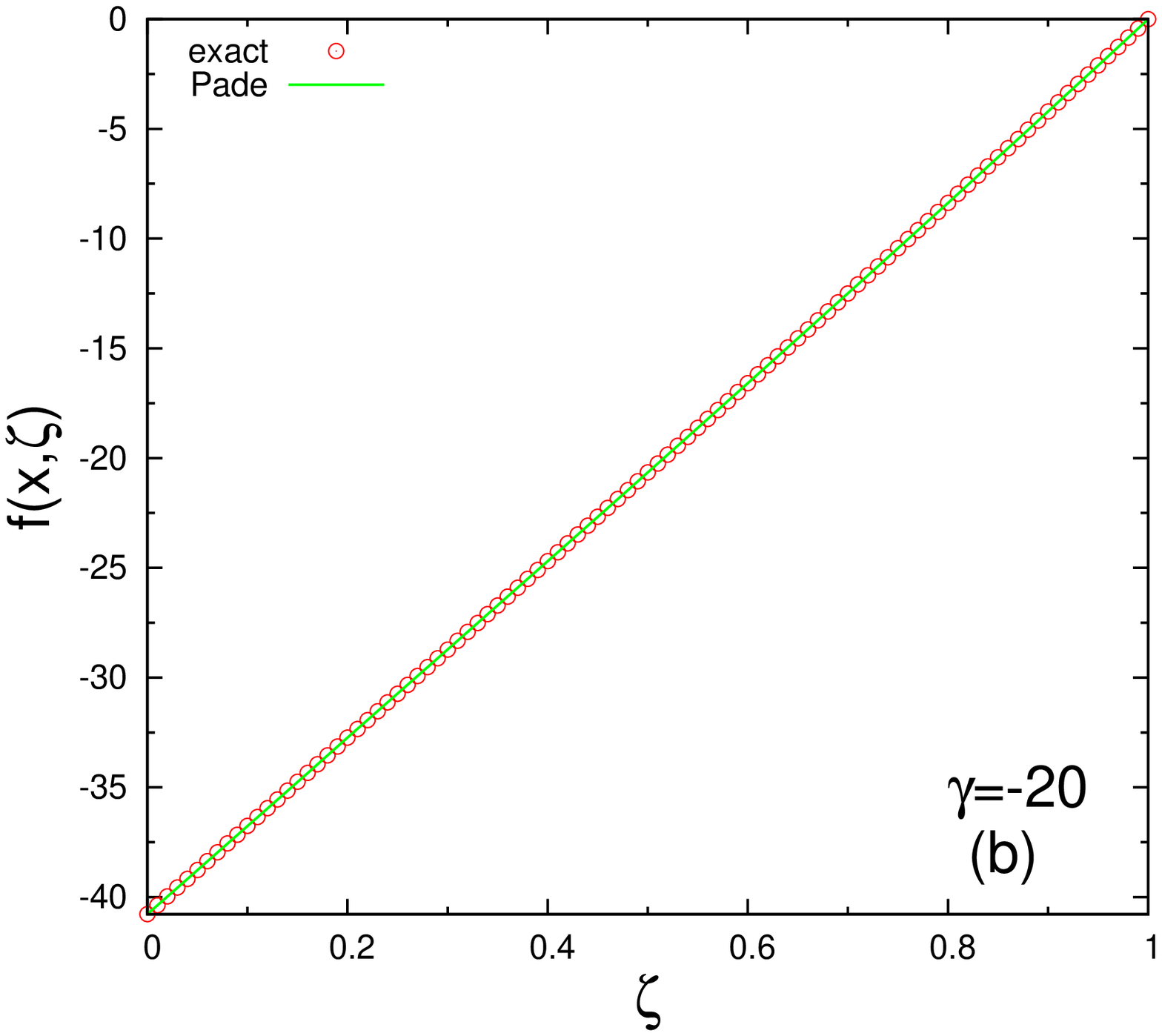}
\caption{(color online) The interaction contribution $f(\zeta,x)=[\varepsilon_{\rm GS}(n,\zeta,\gamma)-\kappa(n,\zeta)]/E_{\rm F}$ to
the ground state energy of the homogeneous Gaudin-Yang model as a
function of the spin polarization $\zeta$ for (a) $\gamma=-2$ and (b) $\gamma=-20$. The
exact result, obtained from the solution of the Bethe-Ansatz
Eqs.~(\ref{eq:rho})-(\ref{eq:normalization_2}), is compared
with the parametrization formula given by Eq.~(\ref{eq:fit_energy}).
\label{fig:one}}
\end{center}
\end{figure}

\begin{figure}
\begin{center}
\includegraphics[width=0.8\linewidth,angle=0]{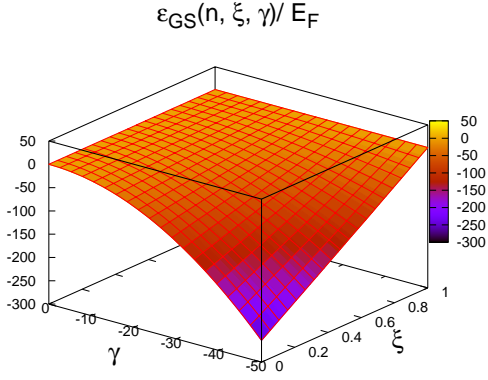}
\caption{(color online) The ground state energy per particle in unit of Fermi energy as functions of $\zeta$ and $\gamma$.}
\label{fig:two}
\end{center}
\end{figure}

For $N$ repulsive fermions interacting {\it via} a $\delta$-type contact, the correlation energy per particle,
and then the GSE, has been parameterized based on Pad$\acute{e}$ approximant in~\cite{Abedinpour,burke_2004,Magyar}. Magyar achieved an
improved parametrization results for the correlation energy by including the higher-order correlation kernel
terms at the high-density limits which is needed for the adiabatic time-dependent DFT approximation~\cite{Magyar}.

For the present system of attractive contact interacting, we solve numerically
a set of coupled Eqs.(\ref{eq:rho})-(\ref{eq:normalization_2}) and find the GSE per particle
$\varepsilon_{\rm GS}(n,\zeta,\gamma) \equiv E/N$ given in Eq.~(\ref{eq:energy_atom})
as a function of $n$, $\zeta$ and $\gamma$. After having $\varepsilon_{\rm GS}(n,\zeta,\gamma)$, we find an accurate parametrization formula to fit $\varepsilon_{\rm GS}(n,\zeta,\gamma)$. Our parametrization formula is
\begin{equation}\label{eq:fit_energy}
\varepsilon_{\rm GS}(n,\zeta,\gamma)=\kappa(n,\zeta)+f(\zeta,x)E_{\rm F},
\end{equation}
where
\begin{equation}
\kappa(n,\zeta)=\frac{\pi^2\hbar^2n^2}{24m}(1+3\zeta^2)
\end{equation}
is the kinetic energy of the noninteracting system per particle and the Fermi  energy
is $E_{\rm F}=\hbar^2 n^2 \pi^2/(8m)$.
Here, $f(\zeta,x)$ can be represented by,
\begin{eqnarray}\label{eq:fit}
f(\zeta,x)&=&[e(x)-1/3]\left\{1+\alpha(x)|\zeta|+\beta(x)\zeta^2\right.\nonumber\\
&-&\left.[1+\alpha(x)+\beta(x)]|\zeta|^3\right\},
\end{eqnarray}
where $x=2\gamma/\pi$, $e(x)$ is given in Ref.~\cite{Gao_PRA06}
very accurately as,
\begin{equation}\label{eq:fit_attractive}
e(x)=\frac{1}{3}-\frac{|x|}{\pi}-\frac{x^2+a_m |x|+b_m}{x^2+c_m |x|+ d_m}\frac{x^2}{4}
\end{equation}
with $a_m=-0.331117$, $b_m=0.458183$, $c_m=a_m+4/\pi$, and $d_m=4a_m/\pi+b_m+16/\pi^2-1$.
Equation~(\ref{eq:fit_attractive}) gives the exact asymptotic behaviors at $x\rightarrow 0^-$
and $x\rightarrow -\infty$~\cite{astrakharchik_2004}.
$\alpha(x)$ and $\beta(x)$ are given by Pad$\acute{e}$ approximant,
\begin{equation}
\left\{
\begin{array}{l}
{\displaystyle \alpha(x)=\frac{x^3+A_\alpha x}{-x^3+B_\alpha x+C_\alpha}}
\vspace{0.1 cm}\\
{\displaystyle \beta(x)=\frac{A_\beta x+B_\beta}{x^3+C_\beta x^2+D_\beta x-B_\beta}}
\end{array}
\right.\,.
\end{equation}
Here $A_\alpha=-0.0652385$, $B_\alpha=-1.87353$, $C_\alpha= 3.08873$,
$A_\beta=3.90515$, $B_\beta=19.4239$, $C_\beta=-4.53457$ and $D_\beta=-7.29826$.

\begin{figure}
\begin{center}
\includegraphics[width=0.8\linewidth,angle=0]{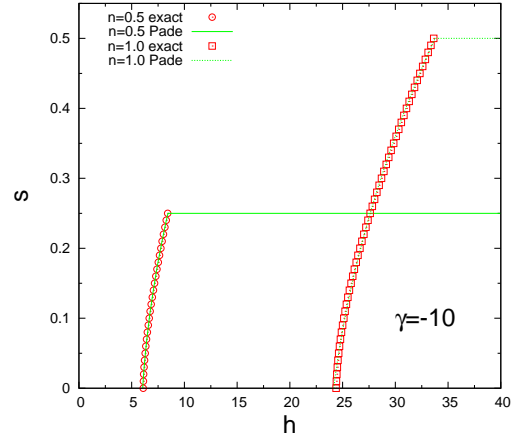}
\caption{(color online) The magnetization (in units of the inverse of arbitrary length $a$)
as a function of magnetic field (in units of $\hbar^2/ma^2$) for different fillings
at $n=0.5$ and $n=1.0$ for given $\gamma=-10$ ($n$ is taken also in units of the inverse of arbitrary length $a$,
in the text we take the same units as described here.).
The exact results from Bethe-Ansatz equations are compared with those derived from the parametrization formula.
\label{fig:three}}
\end{center}
\end{figure}

To assess the validity of our parameterized formula, we show in Fig.~\ref{fig:one} the results
emerging from Eq.~(\ref{eq:fit_energy}) in comparison with the exact Bethe-Ansatz
results for $\gamma=-2$ and $\gamma=-20$ in the weak and strong attractive regime, respectively.
As it is clear from Fig.~\ref{fig:one}, our parametrization formula has an excellent agreement with the exact calculation
and it would be safe to use the parametrization formula for all range of
$n$, $\zeta$ and $\gamma$ values.

Here we would like to deal with some asymptotic behaviors of the simple parametrization formula in Eq.~(\ref{eq:fit_energy}).
First of all, we can simply obtain both unpolarized ($\zeta=0$) and polarized ($\zeta=1$) limits exactly.
Secondly, the weak interacting regime~\cite{Bachelor} is recovered after taking $\gamma\rightarrow 0$ limit
in Eq.~(\ref{eq:fit_energy}) as
\begin{eqnarray}
\varepsilon_{\rm GS}(n,\zeta,\gamma\rightarrow 0)&=&\frac{\hbar^2 \pi^2 n^2}{24m}(1+3\zeta^2)
+\frac{\hbar^2 n^2}{4m}(1-\zeta^2)\gamma \nonumber\\
&&+O(\gamma^2).\nonumber
\end{eqnarray}
The strongly attractive limit from our parametrization formula is
\begin{eqnarray}
\varepsilon_{\rm GS}(n,\zeta,\gamma\rightarrow -\infty)&=&-\frac{\hbar^2 n^2}{8m}(1-\zeta)\gamma^2
 \nonumber\\
&&+\frac{\hbar^2 \pi^2 n^2}{96m}(1+15\zeta)+O(\frac{1}{\gamma})\nonumber
\end{eqnarray}
which has little difference comparing with Batchelor's derivation~\cite{Bachelor}
\begin{eqnarray}\label{Eq:fit_bachelor}
\varepsilon_{\rm GS}(n,\zeta,\gamma\rightarrow -\infty)&=&-\frac{\hbar^2 n^2}{8m}(1-\zeta)\gamma^2
+\frac{\hbar^2 \pi^2 n^2}{96m}\nonumber\\
&&\times(1+15\zeta^3-3\zeta+3\zeta^2)+O(\frac{1}{\gamma}).\nonumber
\end{eqnarray}

For emphasizing these limits, we show $\varepsilon_{\rm GS}
(n,\zeta,\gamma)/E_F$ as functions of
$\zeta$ and $\gamma$ in Fig.~\ref{fig:two}. At $\zeta=0$, $\varepsilon_
{\rm GS}(n,0,\gamma)/E_F$ behaves like $-\gamma$ at weak attractive limit and $-\gamma^2$ at strong attractive one,
while at  $\zeta=1$, $\varepsilon_{\rm GS}(n,1,\gamma)/E_F$ goes as a constant.
The expression at strong interaction limit differs from the corresponding repulsive case
due to the attractive nature of the interaction.
Furthermore $\varepsilon_{\rm GS}(n,\zeta,\gamma)/E_F$ behaves like $\zeta^2$ at weak attractive limit
similar to the corresponding repulsive $1D$ expression, however it behaves like $\zeta$ at strong attractive limit.

Furthermore, we would like to examine some other physical quantities calculated from Eq.~(\ref{eq:fit_energy})
in comparison to those calculated within the exact Bethe-Ansatz equations to obtain more credibility for our parametrization
formula. For this purpose, we take the derivative of the ground state energy with respect to the
magnetization to calculate the magnetic field in equilibrium condition which
is $h(n,s,\gamma)=\partial [n\varepsilon_{\rm GS}(n,\zeta,\gamma)]/\partial s$~\cite{Kocharian}.
The magnetization vanishes when the field becomes smaller than the critical value $h_c$,
a term associated with the spin energy gap in the attractive case (Note that it vanishes in the repulsive case.).
We next display the exact magnetization as a function of the magnetic field
solved numerically from the coupled integro-differential equations based on Eqs.~(\ref{eq:rho})-(\ref{eq:normalization_2})
in comparison to those derived from the parametrization formula.

For given $\gamma=-10$, the exact numerical calculation gives $h_c=6.097$ for $n=0.5$,
while from the parametrization formula we get $h_c=6.049$ for the same density as shown in Fig.~\ref{fig:three}.
Moreover, $h_c=24.388$ for $n=1.0$,
by calculating exact formula and we have $h_c=24.182$ from parametrization formula.
Surprisingly, a finite magnetization appears only if the magnetic field larger than $h_c$.
The magnetization saturates at $s=n/2$ when $h>h_s$.
More precisely, for $\gamma=-10$, the exact solution of $h_s$ is $8.410$ for $n=0.5$,
and $33.643$ for $n=1.0$, which are very close to values obtained from the parametrization
formula that we get $h_s=8.440$ and $33.759$, respectively.

After comparing the exact solutions to the parametrization formula, we convince that
the parametrization formula gives satisfying simulations on the ground state energy and the
derivative quantities like the chemical potential and the magnetic field in a wide range of parameters.

\section{Spin-density-functional theory and Thomas-Fermi approximation}

The ground-state spin densities $n_\sigma(z)$ can be calculated
by spin-density-functional theory (SDFT) solving self-consistently
the Kohn-Sham equations
\begin{equation}\label{eq:kss}
\left[-\frac{\hbar^2}{2m}\frac{\partial^2}{\partial z^2}+V^{(\sigma)}_{\rm KS}[n_\sigma](z)\right]\varphi_{\alpha,\sigma}(z)
=\varepsilon_{\alpha,\sigma}\varphi_{\alpha,\sigma}(z)
\end{equation}
with effective potential $V^{(\sigma)}_{\rm KS}[n_\sigma](z)=V^{(\sigma)}_{\rm
H}[n_\sigma](z)+V^{(\sigma)}_{\rm
xc}[n_\sigma](z)+m\omega^2_{\|}z^2/2$. The ground state density is determined by the
closure
\begin{equation}\label{eq:closure}
n_\sigma(z)=\sum_{\alpha=1}^{\rm occ.}\Gamma^{(\sigma)}_\alpha\left|\varphi_{\alpha,\sigma}(z)\right|^2\,.
\end{equation}
Here the sum runs over the occupied orbitals and the degeneracy
factors $\Gamma^{(\sigma)}_\alpha$ satisfy the sum rule $\sum_\alpha \Gamma^{(\sigma)}_\alpha=N_\sigma$.
The spin-dependent effective Kohn-Sham (KS) potential as usual is composed by  the mean-field term $V^{(\sigma)}_{\rm
H}=g_{\rm 1D}n_{\bar \sigma}(z)$, the exchange-correlation potential defined as the functional
derivative of the exchange-correlation energy $E_{\rm xc}[n_\sigma]$ evaluated at the ground-state density profile,
$V^{(\sigma)}_{\rm xc}=\delta E_{\rm xc}[n_\sigma]/\delta n_\sigma(z)|_{\rm \scriptscriptstyle GS}$,
and the external potential, respectively.

In order to calculate $n_\sigma(z)$, $E_{\rm xc}$ needs to be approximated.
Here we take the local-spin-density approximation (LSDA) which is
known to provide an excellent description of the ground-state properties of a variety of inhomogeneous
systems~\cite{Giuliani_and_Vignale} to calculate. In the following we
employ an LSDA functional based on the parametrization results of Eq.~(\ref{eq:fit_energy}) for the
exchange-correlation potential,
\begin{eqnarray}\label{eq:balda}
E_{\rm xc}[n_\sigma] =\int dz\, n(z)\left.\varepsilon^{\rm hom}_{\rm xc}(n,\zeta,\gamma)
\right|_{n\rightarrow n(z),\zeta\rightarrow \zeta(z)}.
\end{eqnarray}
The exchange-correlation energy $\varepsilon^{\rm hom}_{\rm xc}$ of the homogeneous Gaudin-Yang model is defined as
\begin{eqnarray}\label{eq:xc}
\varepsilon^{\rm hom}_{\rm xc}(n,\zeta,\gamma)=\varepsilon_{\rm GS}(n,\zeta,\gamma)-\kappa(n,\zeta)\nonumber
-\varepsilon_{\rm H}(n,\zeta,\gamma)\,.
\end{eqnarray}
Here $\varepsilon_{\rm H}(n,\zeta,\gamma)=\frac{\hbar^2 n^2}{4m}\gamma (1-\zeta^2)$
comes from the contribution of the Hartree-Fock mean field.

Following the above points, we write down conveniently the Kohn-Sham potential as,
\begin{eqnarray}\label{eq:kspotential}
V^{(\sigma)}_{\rm KS}[n_\sigma](z)
&=&\frac{1}{2}m\omega^2_{\|}z^2+\mu [n,s](z)-\mu_0 [n,s](z)\nonumber\\
&&\pm \frac{1}{2}\left\{h[n,s](z)-h_0[n,s](z)\right\}\,,
\end{eqnarray}
where $\mu [n,s](z)=\partial [n\varepsilon_{\rm GS}(n,\zeta,\gamma)]/\partial n|_{n\rightarrow n(z),s\rightarrow s(z)}$ is
the chemical potential and
$h[n,s](z)=\partial [n\varepsilon_{\rm GS}(n,\zeta,\gamma)]/\partial s|_{n\rightarrow n(z),s\rightarrow s(z)}$
the magnetic field of the homogeneous system evaluated at the local
potentials, respectively. $\mu_0 [n,s](z)$ and $h_0 [n,s](z)$ are the corresponding chemical potential and magnetic field
of the noninteracting system, which are given by
\begin{eqnarray}
\mu_0 [n,s](z)&=&\frac{\hbar^2 \pi^2}{8m}\left[n^2(z)+4s^2(z)\right]\,\,,\nonumber\\
h_0 [n,s](z)&=&\frac{\hbar^2 \pi^2}{m}n(z)s(z).\nonumber
\end{eqnarray}
Here, $n(z)=n_\uparrow(z)+n_\downarrow(z)$ is the total density distribution, and $s(z)=[n_\uparrow(z)-n_\downarrow(z)]/2$ is
the spin density distribution or the local magnetization.

From the discussions in Sec. (\ref{Section:II}), we know that our spin-density-functional
theory with the reference system based on Bethe-ansatz equations has correctly incorporated the
Luther-Emery-liquid nature and the spin energy gap of the homogeneous system~\cite{luther_emery,luttinger_liquids}.
The inhomogeneity nature induced by the harmonic
trap will be incorporated into the exchange-correlation potential through the local spin-density approximation.
Hereafter we refer to this method as BALSDA.

Another simplified method to describe this inhomogeneous system is to resort to a local
density approximation (LDA) also for the noninteracting
kinetic energy function, which is a Thomas-Fermi approximation (TFA) similar to the TFA but implementing the exchange-correlation energy.
We refer to this approach as the TFA in the following.
Within the LDA one assumes that the chemical potential of the trapped system is given by the sum of the spin-dependent local chemical
potential taken to be the chemical potential of the uniform system at the corresponding density, and the external potential.
The ground-state density profile is obtained in the spirit of the local equilibrium condition,
\begin{eqnarray}\label{eq:LDA}
\mu_{\sigma}^{0}=\mu_{\sigma}^{hom}[n_\uparrow,n_\downarrow](z)+m\omega^2_{\|}z^2/2
\end{eqnarray}
derived by directly minimizing the total energy functional of the system. The constants $\mu_{\sigma}^{0}$ are fixed by the
normalization condition $N_\sigma=\int dz n_\sigma(z)$, and $\mu_{\sigma}^{hom}[n_\uparrow,n_\downarrow](z)$, the corresponding
density dependent chemical potentials of the homogeneous system is derived from
the parametrization formula in Eq.~(\ref{eq:fit_energy})
evaluated at the local densities of $n_\uparrow(z)$ and $n_\downarrow(z)$.
Within the LDA, a dimensionless characteristic parameter $\eta=N a^2_{1D}/a^2_{\|}$ can be defined in the harmonic trap.
The systems hold the same density profile if the characteristic parameter is the same irrespective of their different number of
particles and different harmonic oscillators~\cite{astrakharchik_2004}.
The parameter can be also expressed as,
$\eta=4N/\Lambda^2$ with $\Lambda=g_{1D}/(a_{\|}\hbar\omega_{\|})$ if we keep in mind that
$\gamma=-2/(n a_{1D})$. $\eta \gg 1$ corresponds to weak coupling while $\eta\ll 1$
corresponds to the strong interacting regime.

\section{Numerical Results}
\label{sect:numerical_results}
In this section we present the numerical results obtained from the self-consistent solution of
Eqs. (\ref{eq:kss})-(\ref{eq:kspotential})
using our accurate parametrization formula from Pad$\acute{e}$ approximant.
We test the TFA approach and compare its results with our full BALSDA to justify how advance physics we have within BALSDA.
We proceed to illustrate our main numerical results, which are summarized in Figs.~4-5.

\begin{figure}
\begin{center}
\includegraphics[width=0.65\linewidth,angle=0]{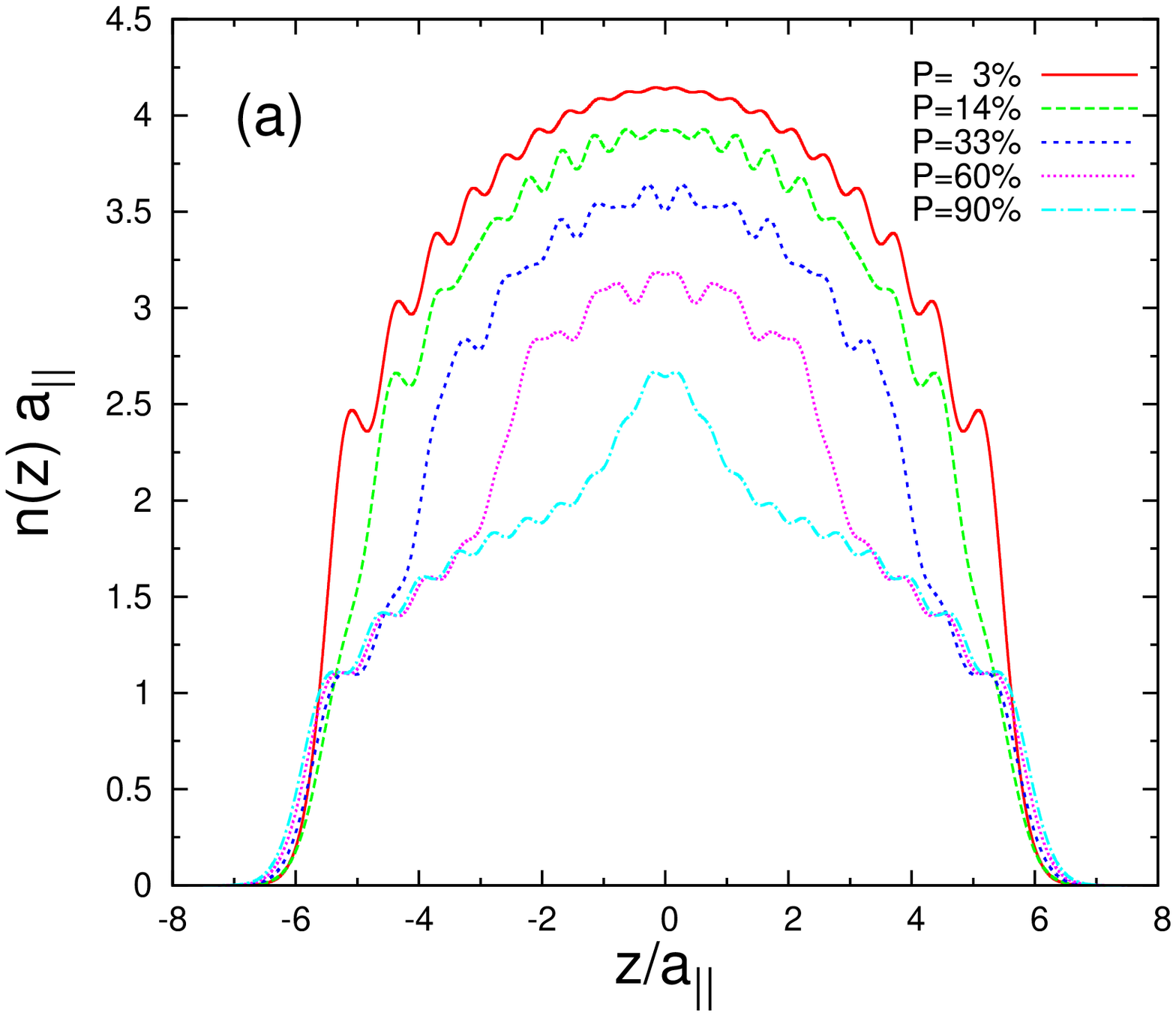}
\includegraphics[width=0.65\linewidth,angle=0]{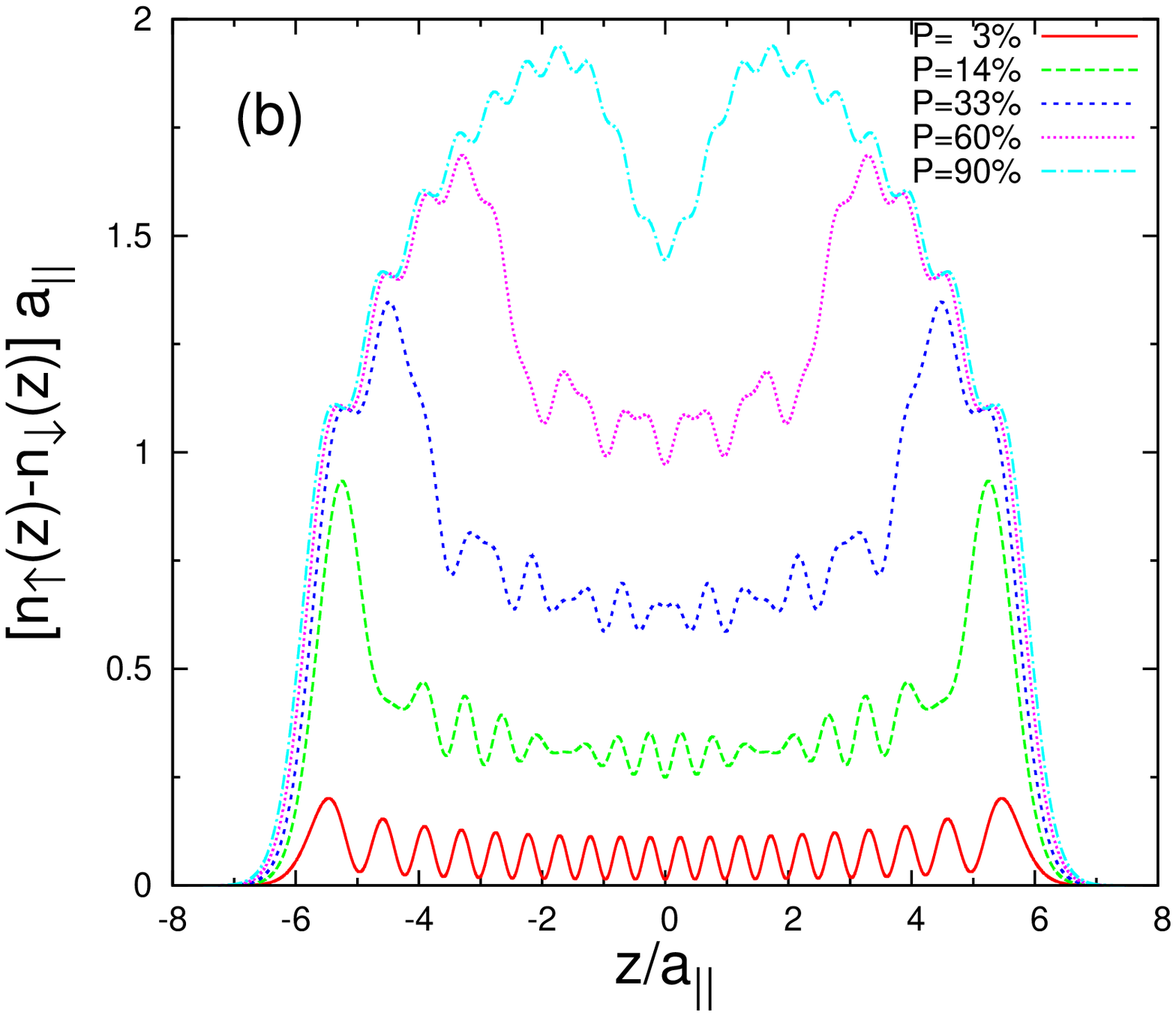}
\caption{(color online) The charge density (a) and spin density (b) profiles as a function of $z$ for
different polarization $P$ at $\Lambda=-2$.
The polarization is increased from $3\%$ to $90\%$ by keeping always a constant number of spin-up atoms, $N_\uparrow=20$ and
decreasing $N_\downarrow$ from 19 to 1.
\label{fig:four}}
\end{center}
\end{figure}

First of all, we show the ground state site occupation for a Fermi gas trapped in a harmonic potential at different polarizations
in Fig.~\ref{fig:four}. Here $\Lambda=-2$ which implies $\eta=N$,
corresponding to weak coupling regime. For the harmonically confined system, we define the average polarization as
$P=(N_\uparrow-N_\downarrow)/(N_\uparrow+N_\downarrow)$,
which is changed keeping always a constant number of spin-up atoms $N_\uparrow$ and decreasing $N_\downarrow$. $P=0$  means
an unpolarized system and $P=100\%$ corresponds to a trivial fully polarized case of $N_\uparrow$ noninteracting fermions.
The density profiles of spin-up and spin-down atoms show $N_\uparrow$ and $N_\downarrow$ maxima, respectively~\cite{Gao_PRA06},
which reflects in the total density profile with $N_\downarrow$ maxima and in the spin one with $N_\downarrow$ minimum.

In Fig.~\ref{fig:four}(a), the total charge density profile is shown as a function of $z$.
The 'hat'-structure is formed in the bulk region by paired minority species
of spin-down and spin-up atoms, manifested by $N_\downarrow$ peaks inside and partially polarized fermions.
Higher polarization gives smaller 'hat'-structure. By increasing the polarization P but
keeping $N_\uparrow$ constant, the range of the charge density becomes a little larger
because the attractive interaction becomes weaker due to the decreasing spin down atoms.
Orso~\cite{Orso} and Liu et al.~\cite{Hu,Hu_2007}, after analyzing the phase diagram and calculating the order parameter,
concluded that the state at the center of trap gives a two-shell structure of the FFLO-type,
which has a spatially oscillatory behavior in the superconducting correlation function and exists in any nonzero polarization
and attractive interaction.
In Fig.~\ref{fig:four}(b), we show the spin density profile defined as the difference between spin-up and spin-down density,
namely the local magnetization as a function of $z$.
By increasing the $P$ value, an oscillating structure of the spin density wave appears in the bulk of the system,
due to the attractive interaction between the majority and minority species.
As a result, a quasi-flat region appears in the bulk part of the spin density profile with coexisted spin-up and spin-down
atoms, surrounded by fully polarized spin-up atoms only.
The core of coexisting spin-up and spin-down atoms becomes smaller and smaller for higher value of $P$.
We would like to point out that the local magnetization distribution as a function of
position is directly measurable through phase-contrast imaging techniques~\cite{Shin}.

\begin{figure}
\begin{center}
\includegraphics[width=0.65\linewidth,angle=0]{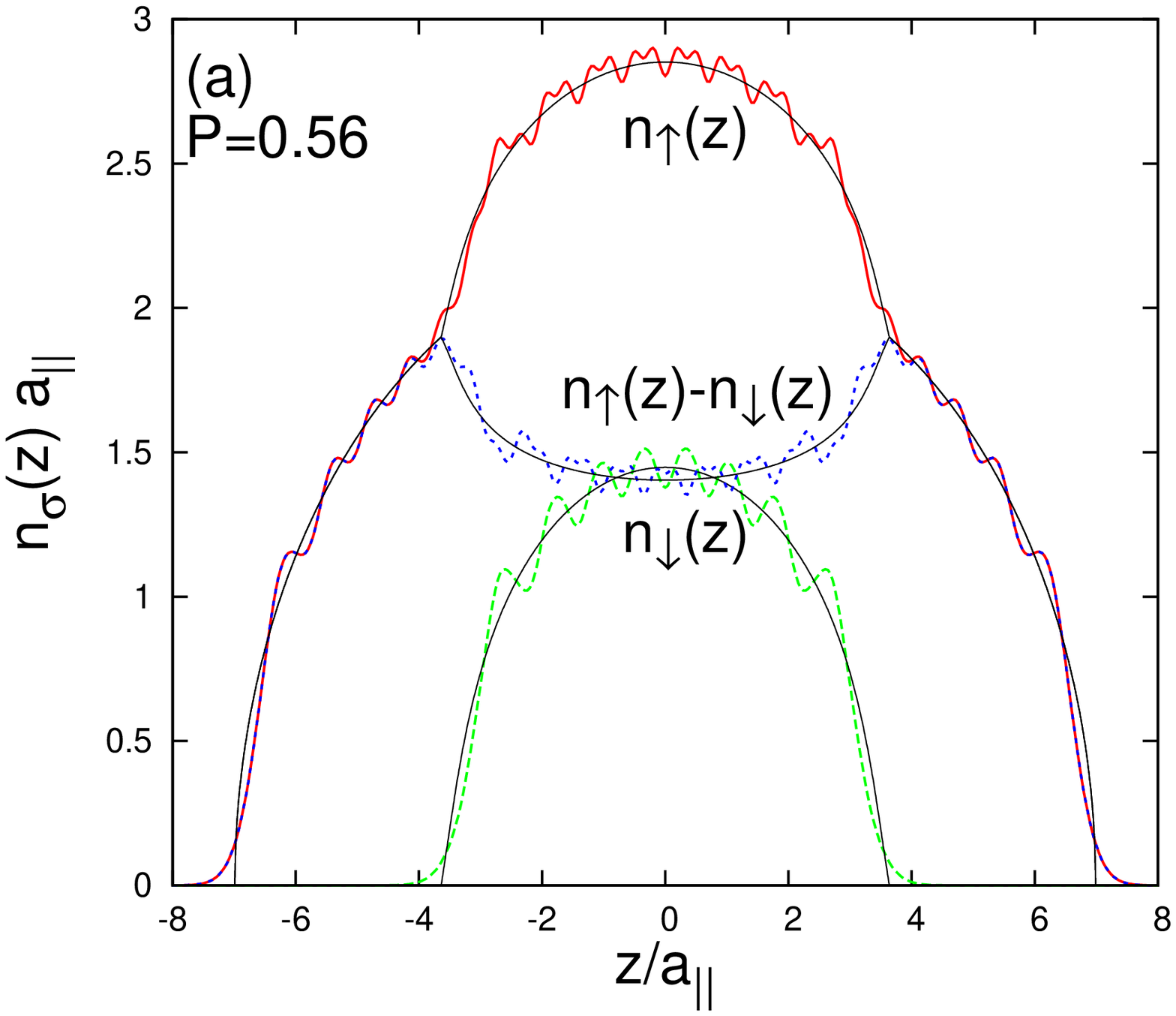}
\includegraphics[width=0.65\linewidth,angle=0]{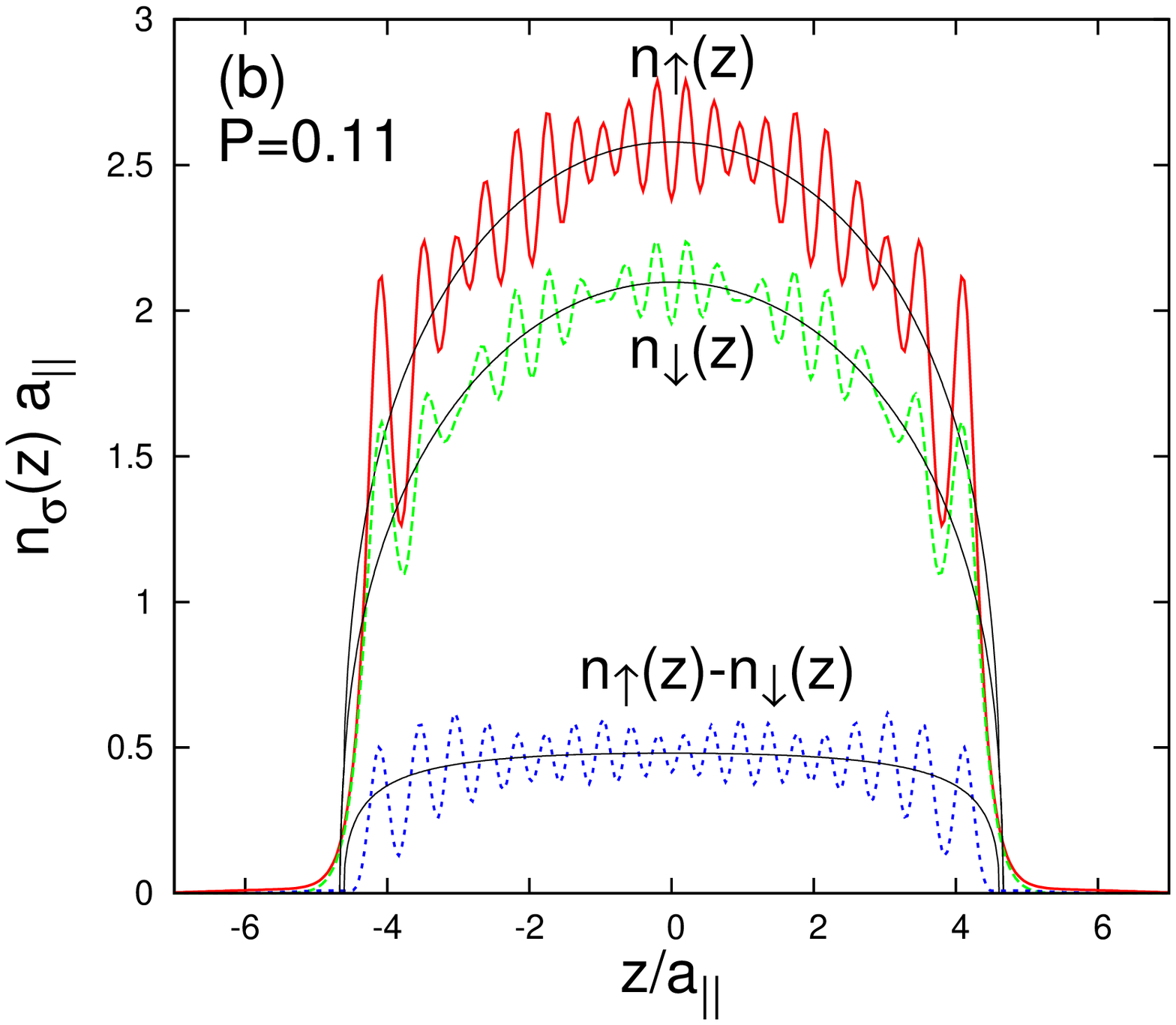}
\includegraphics[width=0.65\linewidth,angle=0]{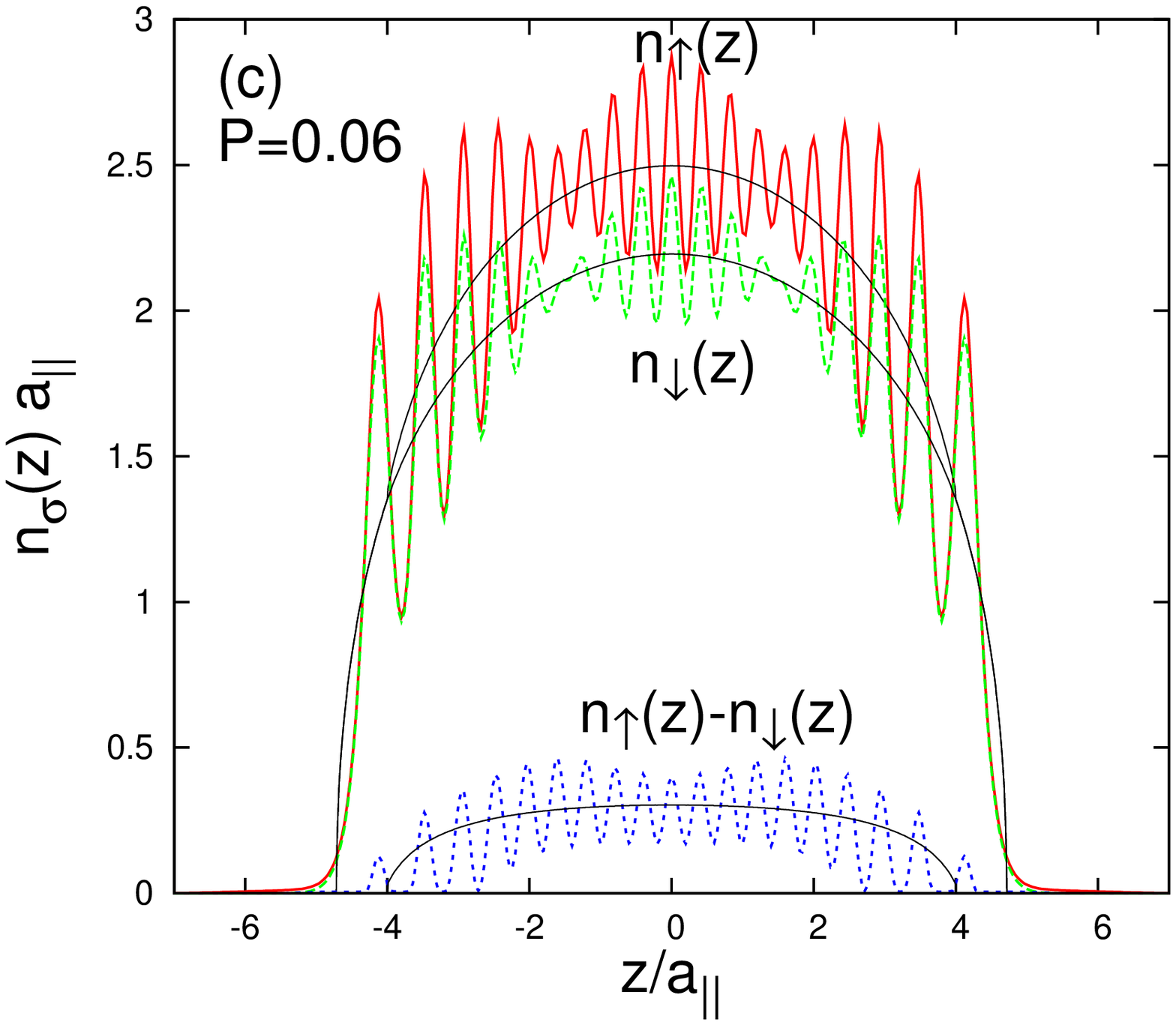}
\caption{(color online) The spin-up, spin-down and spin density profiles of fixed atom number
$N=36$ for $\eta=1$. We show two different results based on the BALSDA (dotted lines)
and the TFA (solid line between the oscillatory BALSDA results).
(a) For $P=0.56$, a two-shell structure is formed with a partially polarized
FFLO-type superfluid of distinctive oscillatory character in the center of the trap and a fully polarized state at the edges.
(b) For $P=0.11$, a two-shell structure is formed with a partially polarized
superfluid in the center of the trap and a fully paired superfluid state at the edges.
(c) For $P=0.06$, a smaller polarization gives larger BCS regime in the wings.
In all the calculations, we show the TFA results give a good overall shape between the oscillating BALSDA ones.
\label{fig:five}}
\end{center}
\end{figure}

In Fig.~\ref{fig:five}, we show the spin-up, spin-down density
profiles, and furthermore, the local magnetization distribution at
$\Lambda=-12$ with fixed total atom number of $N=36$ but varying spin-up
and spin-down atoms. This implies
$\eta=1$, corresponding to an intermediate coupling regime. From Figs.~\ref{fig:five}(a), (b) and (c), it
is clear that the amplitude of the oscillations increases with
decreasing $P$. We present results obtained by the TFA approach together
with BALSDA to justify quantum effects. The TFA results are shown by
solid lines between the oscillating density profiles. For $P=0.56$, see Fig.~\ref{fig:five}(a), the TFA results give a good
representation of the actual density profile, except at the edges of
the cloud. We stress here, without including the exchange-correlation effect, the TFA gives bad overall shape
for the density profiles.

In Fig.~\ref{fig:five}(a), a high polarization case of $P=0.56$ is presented
for the spin-up density, spin-down density and the spin-density profiles.
It is clear that the curve envelop produces an oscillating structure particularly in Figs.~\ref{fig:five}(b) and \ref{fig:five}(c).
A two-shell structure is formed with a partially polarized
FFLO superfluid in the center of the trap~\cite{Hu_2007} and a fully polarized normal state of excess spin-up atoms
at the edges. In Fig.~\ref{fig:five}(b), a polarization case of $P=0.11$ is illustrated with BALSDA together with TFA, where
a different two-shell structure is formed with a partially polarized
superfluid at the trap center and a fully paired BCS superfluid state at the edges, opposite to the 3D counterpart
of the BCS paired phase in the core. In the present system, we observe that there is one critical point which is $P_c=0.12$ between
the two above-mentioned different phase separations which coincides with
the calculations given by Refs.~\cite{Orso,Hu}. We want to stress here, from Fig.~\ref{fig:five}(a) to Fig.~\ref{fig:five}(b),
due to the stronger attractive interaction when the size of the cloud shrinks considerably by decreasing $P$,
that the oscillation structure due to correlation effects become more important and the LDA result becomes worse.
In Fig.~\ref{fig:five}(c), an even smaller polarization case of $P=0.06$ is illustrated with BALSDA against TFA results. The
fully paired superfluid state becomes larger in the wings with such a smaller polarization. The different phases in the two-shell
structure of standard BCS state outside of the center and the FFLO-type state in the bulk,
are proofed to be a smooth second-order transformation~\cite{Hu_2007}.
In all the above calculations with weak and intermediate coupling regimes, we find the TFA results give a good overall shape
to the microscopic calculations based on the BALSDA. But importantly, the TFA could not produce oscillation structure
due to correlation effects in the quantum system.

\section{Conclusions}
\label{sect:conclusions}
In conclusion, we have presented two alternative methods to study the imbalanced
two-component Fermi atomic gases of attractive interactions in Q1D harmonic traps. One is the
spin-density-functional theory based on Bethe-ansatz solutions, which properly incorporates the
Luther-Emery nature and the spin energy gap of the homogeneous system. Another is
the TFA which takes a local density approximation for the noninteracting
kinetic energy but including the exchange-correlation energy. We apply these two methods to the system of finite size
trapped by the harmonic confinement.
At weak coupling strength of any polarization or intermediate coupling of large polarization, a two-shell structure
is obtained with a partially polarized pairs of FFLO-type state in the core
and a fully polarized fermions in the wings. At intermediate coupling regime of small polarization, a two-shell structure
is formed of partially polarized pairs of FFLO-type state in the bulk and a fully paired BCS state at the edges
which is different from the 3D case.
In the current experimental set-up, imbalanced interacting Fermi gases can
be induced by a radio-frequency sweep in the optical lattice system,
and in the on-going experiments, the two-shell structure described here should be observable in the system of thousands of Q1D tubes
formed in the quasi-two-dimensional optical lattices with each of the tube containing up to 100 atoms
measured by absorption imaging techniques,
while the direct test of FFLO states remains as a challenge to experimental community~\cite{Chen}.

It would be of interest to develop the present scheme to study dynamical phenomena in these strongly correlated
gases using time-dependent
DFT and/or current-DFT, instead of resorting to the inhomogeneous Tomonaga-Luttinger liquid model. From a more
formal DFT viewpoint, a functional
better than the one presented in Eq.~(\ref{eq:balda}) is desirable and necessary to deal with
the strong coupling regime.

\acknowledgments
G. Xianlong was supported by NSF of China under Grant NO. 10704066.
We would like to thank Hui Hu, Marco Polini and  Saeed Abedinpour for many useful discussions.

\end{document}